\documentclass{aa}
\usepackage{graphics}

\begin{document}                                          
\thesaurus{08.02.3, 08.12.2, 05.01.1, 03.20.4, 08.02.6, 08.09.2 Sirius}
\title{Search for companions around Sirius\thanks{Based on data collected 
at the Pic du Midi-Toulouse Observatory (France)}}
\author{J.M. Bonnet-Bidaud\inst{1} 
\and F. Colas\inst{2}
\and J. Lecacheux\inst{3}} 
\offprints{J.M. Bonnet-Bidaud}
\mail{bobi@discovery.saclay.cea.fr}
\institute
{Service d'Astrophysique, DSM/DAPNIA/SAp, CE Saclay, 
F-91191 Gif sur Yvette Cedex, France
\and
Institut de M\'{e}canique C\'{e}leste, 77 Av. Denfert Rochereau, 75014-Paris, France
\and
DESPA, Observatoire de Paris-Meudon, 5 Pl Jules Janssen, 92195-Meudon, France}
\date{Received date: ; accepted date: }
\maketitle

\begin{abstract}
Since the discovery of Sirius-B about 130 yr ago, 
there have been several claims of a possible second companion
around the brightest star Sirius-A. Such a companion could,
in particular, be responsible of the suspected colour change of the star, 
now strongly suggested from two independent historical sources. 
We reported here on a new observation of the sky region around Sirius, 
to search for such a companion, using a coronographic device.
 
By comparison of the new stellar field with a similar image obtained 
by us  $\sim$13 yr ago and using the Sirius proper motion,
we are able to eliminate the most obvious companion candidates 
down to a magnitude  m$_v$$\sim$17 in a field from 30 arcsec to 2.5 arcmin
of the central star. None of the visible stars appears consistent in
magnitude and colours with what expected from current theoretical models
and observations of low-mass stars.

From the study of the same field, it is also shown that  the Sirius companion, 
consistently reported  by observers during the years 1920-1930, is most probably
an unrelated m$_g$$\simeq$12 background star, now $\sim$ 1 arcmin away but 
located precisely on the Sirius proper motion trajectory. The closest
apparent conjunction with Sirius  was realized in 1937 with 
a minimum angular distance of 6.9 arcsec, of the same order than the 
Sirius A-B binary separation.

The reported observations do not eliminate the possibility of a second
 companion but now confined the search to the more central 30 arcsec region 
around Sirius. In particular, the existence of a long period 
companion cannot definitively be ruled out since the arbitrary orientation 
of the orbit can yield an observed projected position on sky inside this
more central region.
\end{abstract}

\section{Introduction}
Sirius is the brightest star in the sky. Outshining the sun by a factor
 25 in absolute luminosity, it displays at a distance of  2.64 pc
an apparent visual brightness that exceeds by a factor 2
the one of the second brightest star, Canopus (Gatewood \& Gatewood 1978). 
Due to its fast relative motion through the sky with a radial 
motion of 1.6 AU/yr and a proper motion projected on sky of 1$\degr$ in 
about 2700 yr, it will not always be the dominant star in the sky though 
this status is already true for 90~000 yr and will remain valid 
for at least 210~000 more years (Tomkin 1998).
Its velocity will bring Sirius at a minimal distance from the Sun of 2.3 pc
in about 65 000 yr (Garcia-Sanchez et al. 1998).
It is from the study of this apparent fast motion through the sky, 
first noted by Halley, that the existence of the famous white dwarf 
companion, Sirius -B, has been predicted by Bessel (1844), making
of Sirius-B one of the first example of dark matter body whose existence
has been inferred from its gravitational action only. 
Sirius-B was finally discovered in the early days 
when the first modern refractor was put into service in Massachusetts (USA)
(for an historical review see Brecher 1979, Hetherington 1980, Wesemael 
and Fontaine 1982). 
After the discovery of Sirius-B, the binary system  has been  
extensively observed first visually (Volet 1932, Zagar 1932), 
then by photographic techniques (Lindenblad 1970) to accurately determine 
the orbit.
Since the discovery, the orbit has now been covered  a little more than twice 
so that the orbital elements are now relatively well established 
(Gatewood \& Gatewood 1978, Benest \& Duvent 1995). The distance
and proper motion of the system were also recently refined by 
the Hipparcos satellite observations (ESA 1997).

The existence of possible additional companion(s) to Sirius-A is still 
however an open question.
The star itself is well known but the sky region immediately
around it is poorly explored. Observations are  strongly affected by
the central star strong contamination by diffusion,  
and a direct evidence of a companion is still extremely difficult 
to establish even with modern techniques.

There have been several repetitive claims that a visual (m$_v$=12 ?) 
companion have been detected during the years 1920-1930 
(Fox 1925, Innes 1929, van den Bos 1929, see Baize 1931 for a review).
Studies of Sirius orbital perturbations (Volet 1932, Zagar 1932), 
as well as more recent detailed analysis of the orbital residuals 
(Benest \& Duvent 1995), have also led to a persistent claim of a possible 
6 years periodicity. This points towards a very low mass companion orbiting most
probably Sirius-A with a suggested separation of the order of 2-3 arcsec.

Independently, there is increasing indirect evidences, 
that a second companion with a long period could be at the origin 
of a colour change of the bright star Sirius in historical times. 
The tidal interaction of this low-mass companion during the close-by encounter 
at periastron could cause enough matter to be expelled and produce an 
appreciable transient reddening of Sirius (Bonnet-Bidaud \& Gry 1991).
This event, which could have happened a few centuries before modern era, 
is now suggested from independent greek and chinese historical records
 (Gry \& Bonnet-Bidaud 1990, Bonnet-Bidaud \& Gry 1991, 1992).
Such companion could be expected to reach a separation of the order
of 60 arcsec.

From previous optical observations that we performed at ESO (Chile) in 1985, 
an image has been obtained around Sirius, showing for the first time
all stars from 30" to 2.5' away from the binary, down to a magnitude 
m$_v$$\sim$17. 
From this detailed image of the field, 
several possible candidates matching the apparent magnitude 
and positions expected from a long period companion have been singled out 
(Bonnet-Bidaud \& Gry 1991).

We present here the follow-up of this program for the search of new 
companion(s) of Sirius-A. From the use of the two images obtained 
at a 13 years interval, a definitive identification test can be carried out,
using the unusual large proper motion of Sirius-A.

\section{Observations}

Photometric observations of the Sirius field were  carried out 
in good seeing conditions on January 18-24th, 1999 using a 288x384 pixels 
CCD camera at the  Nasmyth  secondary  focus of the 1 meter telescope 
in the Observatoire of Pic du Midi-Toulouse (OPMT, France).
To suppress most of the contamination from the bright central Sirius, a
special coronographic device was designed. A small circular mask
was fixed on a thin metallic wire and mounted on a two-dimension
moving plate to allow for its precise positioning in the optical path.
This mask was introduced at the  intermediate focus of a f/D=6 focal reducer.
The resulting field was 4.8x3.6 arcmin centered at Sirius with a
pixel size of 0.75 arcsec and a reconstructed mask size of $\sim$ 22 arcsec
in diameter.
Different attempts were made to further reduce the strong cross-like spikes
originating from the diffusion by the supports of the secondary mirror.
For this, a  plywood cache with four-circular apertures was set up at the entrance 
of the telescope.
Through reducing the first order high spatial frequencies, 
this cache was found to introduced an additional diffusion in the central region 
of the field so images both with and without this cache were obtained.\\
Individual images were taken with 2-3 min. exposures using a Gz Gunn filter (Schott RG800)
which, combined with the CCD response, yields a (830-950)nm response.
Images were co-added to produced average images with equivalent 
exposures from 30 to 80 min. The occulting mask was shifted at different positions 
to allow the coverage of a wider field around Sirius.\\

\section{Analysis and results}
The photometric reduction was performed with standard techniques, using the
stars Landolt 653 and 667 as comparisons.
After offset and flat-field corrections, the images were further 
corrected from the residual Sirius diffusion by subtracting 
a background filtered by a radial gradient method.
The resulting Gz Gunn image is shown in Figure 1. 
The image is a composite of three exposures of 81min,
60min and 30 min. obtained with Sirius at center and displaced east and west
respectively.

\begin{figure*}
\resizebox{\hsize}{!}{\includegraphics{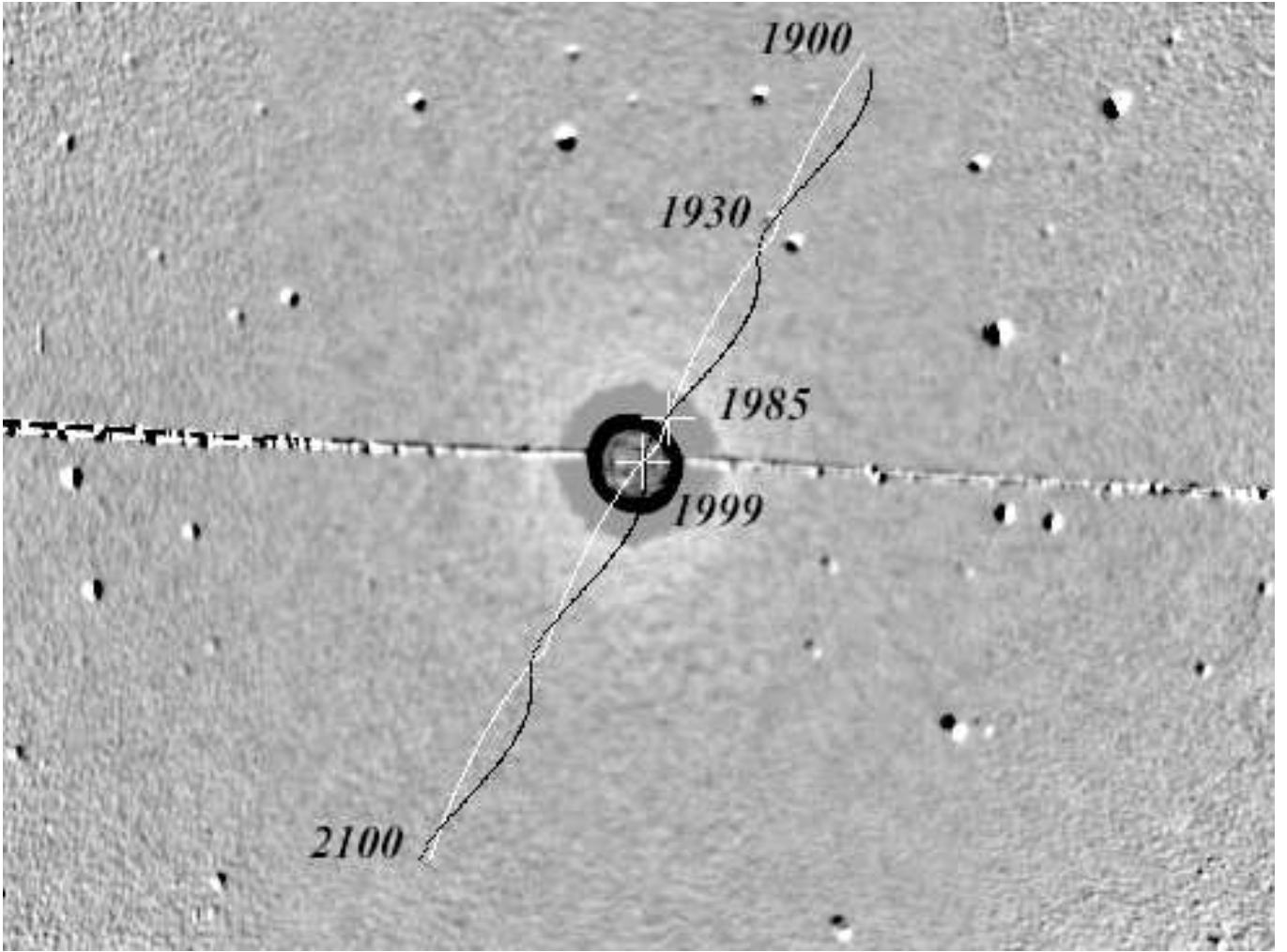}}
\caption[ ]{The (4.5x6.0 arcmin) field centered on Sirius, obtained 
through a Gz Gunn (830-950nm) filter (North is up, East is right).  
The central mask has a reconstructed diameter
size of $\sim$ 22 arcsec. The residual diffusion background have been removed 
by a radial gradient filter, which introduced the visible artifact asymmetry 
in the stellar images. 
Also drawn is the Sirius-A
trajectory (white line), including the proper motion and the orbital 
influence of Sirius-B (black line).
The positions of Sirius-A in 1985 and 1999 are marked by crosses.}
\end{figure*}

\begin{figure}
\resizebox{\hsize}{!}{\includegraphics{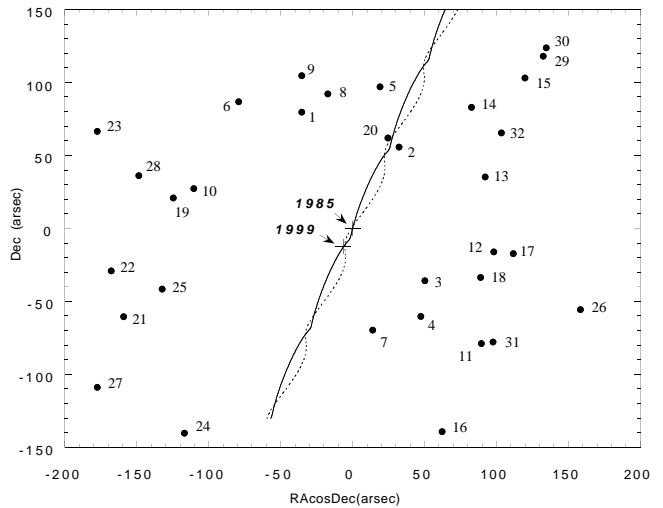}}
\caption[ ]{Positions and identifications of the stars in the Sirius field.
The positions of Sirius-A (full line) and B (dotted line) are also shown along time.
Note the close approach between Sirius and star 2 and 20 in the past, around year 1930.}
\end{figure}

A total of $\sim$30 stars 
is detected in the field down to a magnitude of m$_{g}$$\sim$16 for a region
at a distance greater than 30 arcsec from the center of the mask.
By computing the center of the diffusion profile through the image, we 
checked that the center of the mask corresponds to the Sirius position
with an accuracy better than 0.5 arcsec.
The positions of the different stars in the field were then measured 
by means of gaussian fits to the stellar images, 
yielding an accuracy in the relative positions with respect to Sirius 
better than 0.05 arcsec, except possibly for the few stars falling near regions of 
increased diffusion. Measurements were done both on raw and 
background-subtracted images. 
The positions and labels of the different  stars are shown in Figure 2 with
their coordinates and magnitudes reported in Table 1.
The accuracy on magnitudes is not better than 0.1 mag due to the high background subtracted.
Absolute J2000 coordinates were computed by mean of the astrometric reduction program 
developed at the IMC/Bureau des Longitudes, using the J2000 position
of Sirius-A at 1985.9589 (  $\alpha$=~6h 45m 9.36s, $\delta$=~-16d 42m 43.55s) as
a reference. The relative accuracy is better than 0.1 arcsec but the absolute accuracy
is only ($\sim$0.5 arcsec), dominated by the uncertainty in the Sirius position in the field.

\begin{table}
\caption[ ]{Star magnitudes and positions}
\begin{flushleft}
\begin{tabular}{lrrr}
\hline
Star & mGz  & 
    \multicolumn{2}{c}{Absolute position}  \\
     &    &   \multicolumn{1}{c}{$\alpha$ }  &   \multicolumn{1}{c}{$\delta$ } \\
\hline
 \multicolumn{2}{l}{Sirius} & 6 45 09.35 & -16 42 43.55  \\
 1 & 11.74 &  6  45  06.87  & -16 41 23.85 \\
 2 & 12.63 &  6  45  11.61  & -16 41 47.85 \\
 3 & 13.03 &  6  45  12.87  & -16 43 19.35 \\
 4 & 13.76 &  6  45  12.68  & -16 43 43.95 \\
 5 & 14.05 &  6  45  10.69  & -16 41 06.45 \\
 6 & 13.66 &  6  45  03.82  & -16 41 16.85 \\
 7 & 13.71 &  6  45  10.34  & -16 43 53.25 \\
 8 & 15.41 &  6  45  08.16  & -16 41 11.45 \\
 9 & 15.00 &  6  45  06.88  & -16 40 58.95 \\
10 & 13.49 &  6  45  01.64  & -16 42 16.30  \\
11 & 12.05 &  6  45  15.63  & -16 44 02.43 \\
12 & 12.46 &  6  45  16.23  & -16 42 59.62 \\
13 & 10.83 &  6  45  15.82  & -16 42 08.37 \\
14 & 14.08 &  6  45  15.15  & -16 41 20.67  \\
15 & 12.02 &  6  45  17.76  & -16 41 00.48  \\
16 & 13.89 &  6  45  13.72  & -16 45 02.78  \\
17 & 13.46 &  6  45  17.19  & -16 43 00.88  \\
18 & 13.83 &  6  45  15.59  & -16 43 17.19  \\
19 & 14.87 &  6  45  00.63  & -16 42 22.73 \\
20 & 14.69 &  6  45  11.09  & -16 41 41.58 \\
21 & 13.66 &  6  44  58.22  & -16 43 44.12 \\
22 & 13.20 &  6  44  57.62  & -16 43 12.56  \\
23 & 15.37 &  6  44  56.94  & -16 41 37.07  \\
24 & 15.22 &  6  45  01.19  & -16 45 03.81  \\
25 & 14.58 &  6  45  00.09  & -16 43 25.11  \\
26 & 14.90 &  6  45  20.44  & -16 43 39.23 \\
27 & 15.39 &  6  44  56.94  & -16 44 32.52  \\
28 & 15.88 &  6  44  58.95  & -16 42 07.34  \\
29 & 16.56 &  6  45  18.64  & -16 40 45.51 \\
30 & 15.46 &  6  45  18.78  & -16 40 39.80 \\
31 & 15.36 &  6  45  16.20  & -16 44 01.48 \\
32 & 15.98 &  6  45  16.60  & -16 41 38.10 \\
\hline
\multicolumn{4}{l}{* for accuracy see text}\\
\end{tabular}
\end{flushleft}
\end{table}

This 1999 Sirius field was compared to a similar image obtained 
by us in 1985 with a different telescope. Within this time interval, 
Sirius is expected to have moved through the sky by more than 17 arcsec, 
so that any star dynamically linked to it will also show a significant 
proper motion.
To judge of possible motion of stars in the field, the star positions
of the 1999 image were cross-correlated with the corresponding positions
measured on the 1985 image (see table 1 of Bonnet-Bidaud \& Gry 1991). 
Table 2 lists the corresponding residuals after transformation. 
The mean measured residuals are 0.18 and 0.14 arcsec respectively for the right ascension 
and declination. 
No significant changes are apparent.
The maximum difference is seen for star 8 (0.55 arcsec),
still comparable to the accuracy of the measurements. Some of this deviation may
arise from a small proper motion of the star itself.

\begin{table}
\caption[ ]{Star residuals(*)}
\begin{flushleft}
\begin{tabular}{rrr}
\hline
 Star &  \multicolumn{1}{r}{$\alpha$cos$\delta$(")} & \multicolumn{1}{r}{$\delta$(")} \\
\hline
 1 & +0.03  & +0.00\\
 2 & -0.03  & -0.11\\
 3 & -0.03  & +0.01\\
 4 & -0.05  & -0.10\\
 5 & -0.05  & -0.10\\
 6 & +0.29  & -0.00\\
 7 & +0.04  & -0.17\\
 8 & +0.49  & +0.25\\
 9 & -0.41  & -0.07\\
\hline
\multicolumn{3}{l}{* with respect to 1985, see text}\\
\end{tabular}
\end{flushleft}
\end{table}

By comparison, the position of Sirius was also determined with respect to this
cross-correlated stellar frame. The values are given in Table 3 with a shift
of -(6.6$\pm$0.5) and -(12.0$\pm$0.7) arcsec, respectively in right ascension 
and declination.
Based on the proper motion accurately determined from the recent
Hipparcos data (ESA 1997), the expected shift for the center of gravity of 
Sirius A-B is listed in Table 3 together with 
the corresponding position of Sirius-A,
computed using the orbital ephemeris of Gatewood \& Gatewood (1978).
The predicted change in Sirius-A position is -(6.32$\pm$0.02) in right ascension
and -(12.51$\pm$0.02) arcsec in declination, 
in correct agreement with the measurements, 
if one allows for the uncertainty introduced by the mask.\\

\begin{table}
\caption[ ]{Sirius positions}
\begin{flushleft}
\begin{tabular}{llll}
\hline
 & Date &  \multicolumn{2}{c}{Relative position}\\
 &      &  \multicolumn{1}{c}{$\alpha$cos$\delta$(")} & \multicolumn{1}{c}{$\delta$(")} \\
\hline
                    & 1985.959 &  0.00 (reference) & 0.0 (reference)\\
A (measured) & 1999.052 & -6.65 (0.50) & -12.03 (0.70) \\
G (predicted)& 1999.052 & -7.150 (0.017) & -16.014 (0.016)\\
A (predicted)& 1999.052 & -6.320 (0.020) & -12.510 (0.019)\\
\hline
\end{tabular}
\end{flushleft}
\end{table}

\section{Discussion}
\subsection{Historical claim of a companion}
The influence of the unusual large proper motion of Sirius have often been 
neglected when discussing the properties of the system. The image obtained
here allows the positions of the closest stars to the system to be confirmed.
With the most recent proper motion measurements by Hipparcos, the precise
position of Sirius-A within this stellar field can therefore be extrapolated
backward in time. The exact trajectory of the star, including both the proper 
motion and the orbital influence of Sirius-B is shown in Figure 1 and 2.
The star path is very close to the bright (m$_{g}$$\sim$12) star 2 and 
to the fainter star 20 of our list.
The separation between Sirius and the stars 2 and 20 are computed along time 
in Figure 3, together with the Sirius A-B separation for comparison. 

\begin{figure}
\resizebox{\hsize}{!}{\includegraphics{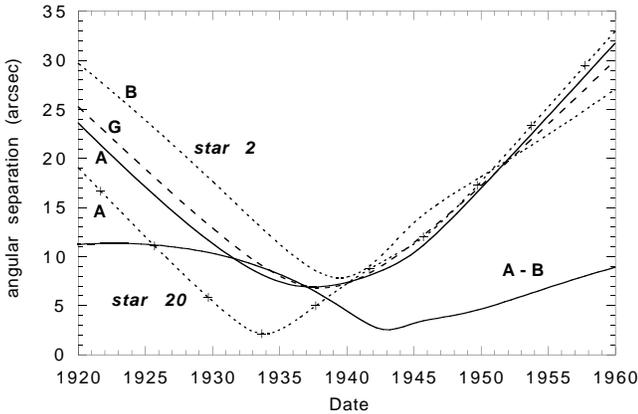}}
\caption[ ]{The apparent separation projected on the sky along time 
between Sirius and star 2 (full line : Sirius-A, thin-dotted line : Sirius-B, 
dotted line : Sirius center of gravity) and 20 (thin-dotted-cross :Sirius-A). 
Also shown is the Sirius A-B separation (lower curve)}
\end{figure}

The bright star 2 was less than 20 arcsec away from
Sirius-A between 1923 and 1951 and within 10 arcsec in the 1932-1944 interval.
The closest approach is reached in october 1937 with a separation of 6.9 arcsec,
of the same order than the Sirius A-B separation. The fainter star 20 is
even closer with a minimum at $\sim$2 arcsec around november 1934.\\
Interestingly enough, the presence of an additional star around Sirius
has been repeatedly claimed by visual observers, more particularly in
this period 1920-1930 when the system was extensively observed
to refine the ephemeris near maximum elongation (see Fox 1925, 
van den Bos 1929, Innes 1929).
Such claim was later found unsupported from  photographic observations 
(Lindenblad 1973), casting  doubts on the quality of these past optical 
observations.
The companion was quoted to be a relatively bright (m$_{v}$$\sim$12) star. 
It is clear that the reported companion was either star 2 or 20 but most likely
the brightest background star 2 which was at a very close distance at the same epoch 
and bright enough to be duly observed as such by the careful optical observers.

\subsection{Characteristics of a Sirius low-mass companion}
From the present observations, none of the brightest stars in the Sirius 
close field appears to have a proper motion comparable to Sirius A-B.
From the constraints on the orbital period and periastron distance, the 
possible orbit of a long period companion is a highly eccentric ({\it e} $\geq$ 0.9)
orbit with a period {\it P} $\geq$ 2000 yr and a semi-major axis {\it a} $\geq$ 230 AU
(Bonnet-Bidaud \& Gry 1991). For an orbit with {\it e}=0.9 and {\it a}=230 AU,
the maximum orbital velocity 
will be  {\it V} = 15 km.s$^{-1}$ at periastron while it will
drop to  0.8 km.s$^{-1}$ at apastron, with a maximum separation of 160 arcsec. 
Outside the most inner part of the orbit, at a distance greater than 30 arcsec to
the central binary, the upper limit on the companion 
orbital velocity will be $\leq$7.7 km.s$^{-1}$ for {\it a}$\geq$ 230 AU, 
therefore significantly less than the Sirius tangential 
velocity (16.795 km.s$^{-1}$). 
In the explored region, at a distance greater 
than $\sim$ 30 arcsec, one would then expect the proper motion of a
companion to be closely comparable to the one of Sirius.

The investigation of the stellar field around Sirius is now complete down
to a magnitude $\sim$17 in V and $\sim$16 in Gz, for distances between 30" and 160".
The absence of a significant proper motion among the bright stars in the Sirius field
rules out the existence of a plausible companion at a today projected distance greater 
than $\sim$30 arcsec from Sirius-A.
 
Since our last 1985 observations, considerable progress has been made in
the understanding of low (0.1-0.5M$_{\sun}$) and very low mass 
(0.07-0.1M$_{\sun}$) stars, 
down to the brown dwarf limit, both theoretically (see Baraffe et al. 1998 and
references therein) and observationally (see Leggett et al. 1998).
The characteristics of a low mass companion at
the Sirius distance can therefore be evaluated more precisely.

Figure 3 shows a Mv-(V-I) color-magnitude diagram for a sample of 
observed small mass red dwarfs from the compilation
by Leggett et al. (1998), in which the stars have been selected for 
their solar-metallic abundances [m/H]=0 and a mass $\leq$0.12M$_{\sun}$.
The selection also includes the newly discovered brown dwarf, Kelu-1,
for which a distance of 10pc have been estimated (Ruiz et al. 1997).
Also shown is the most recent theoretical model for solar-metallicity
low-mass stars,  based on non-grey atmospheres (Baraffe et al. 1998)
with masses indicated. As noted by the authors themselves,
as a probable consequence of an underestimate of opacity, 
this best present model 
is however known to underestimate the V-I
colour by about 0.5 mag for M $\leq$0.5M$_{\sun}$ as visible in Fig.4.
Excluding brown dwarfs, the expected apparent visual magnitude of a 
(0.08-0.12M$_{\sun}$) red dwarf at the Sirius distance (2.64pc) is in the range
m$_{v}$=12-17 with corresponding colour (V-I)=3.0-4.5.

The measured stars in the Sirius close field are also shown in Figure 3 with
 absolute magnitudes as if they were at the Sirius distance. Obviously,
all candidates appear severely underluminous for their colours. The
reddest candidates  (star 3 and 4) have colour marginally consistent with
low mass ($\sim$0.1-0.15M$_{\sun}$) stars but their magnitudes will put them at
distance respectively (50-60)pc and (40-50)pc 
if they are true standard main-sequence stars.

\begin{figure}
\resizebox{\hsize}{!}{\includegraphics{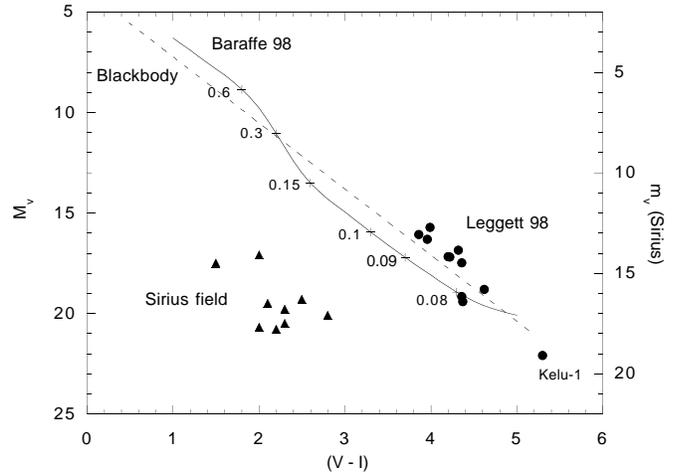}}
\caption[ ]{Mv-(V-I) expected colour-magnitude diagram for potential 
companions of Sirius, down to the brown dwarf limit. Ordinate scales show both 
absolute magnitude (Mv) and apparent magnitude scaled at the Sirius distance (m$_v$).
Full line : theoretical model (Baraffe et al. 1998) with stellar 
masses indicated; dotted line : equivalent
colours of a blackbody atmosphere; full circles : observed low-mass red dwarfs
(Leggett et al. 1998). Also shown are the measured stars in Sirius field
(full triangles). All these stars appear too severely 
underluminous to be consistent with the assumed Sirius distance.}
\end{figure}

We note that the magnitude limit of m$_v$$\sim$17 allows to exclude all possible 
main sequence stars above the brown dwarf limit but the faintest brown dwarfs 
such as Kelu-1 at a magnitude m$_v$=18-19 will be still undetectable .

\section{Conclusion}

The careful study of the Sirius stellar field at a several years interval
allows to exclude the presence of a main sequence low mass 
($\geq$0.08M$_{\sun}$) companion at a separation greater
than $\sim$30 arcsec of the Sirius A-B binary. Only the most extreme case 
of a brown dwarf star comparable to Kelu-1 can be still undetectable. 
The most central region around the binary is however still unexplored. 
Even in the case of a long period companion, 
a significant probability exists that an eccentric orbit will
bring the low-mass star at a projected distance closer than 30 arcsec, 
according to the orientation of the orbit in space. 
As a result of a numerical simulation with random values for the relevant 
orbital parameters, the probability is computed to be $\sim$10$^{-2}$ in the 
less favourable case of an object now at the maximal apastron distance for
a representative orbit with {\it a}=230AU and {\it e}=0.9.
This inner region could also harbour a short orbit companion which
may be responsible for the suspected cyclic residual orbital variations.
The orbital velocity of a close companion could prevent to use the proper
motion discrimination such as used in this paper but 
the characteristics of the low and very low mass stars (luminosity and colours)
 are now sufficiently known to allow a secure identification among 
background stars. 
The Sirius pair is unique in showing for Sirius-A the earliest stellar type 
among white dwarf companions together with a very  unusual high mass among
white dwarfs for Sirius-B. Those peculiarities are still largely unexplained 
and could possibly be in relation with the evolution of a more complex system 
including a yet undetected third star.
It is clear that the study of the most inner part of the sky region 
around Sirius
would require the use of the infrared bands to take advantage of the reduced
contribution of Sirius A-B  together with the increase flux
expected from low mass stars in this range.  These wavelengths also benefit
of the new possibilities of adaptive optics now available.

\medskip
\it Acknowledgments. \rm We are very grateful to J.M. Abaddie and the whole OPMT technical team
for their efficient help in the short term planning of these observations.
We also wish to thank J\'{e}r\^{o}me Blumberg for the realisation of a video film describing
the set-up and the realtime observations. The film "Le compagnon de Sirius" is  available 
at the CNRS Image Media, 27 rue Paul Bert, 94200 Ivry, France

\end{document}